\documentclass[pra,aps,showpacs,showkeys,groupedaddress,superscriptaddress,twocolumn]{revtex4-1}

\usepackage[utf8x]{inputenc}
\usepackage{color}
\usepackage{bbm} 
\usepackage{amsfonts,amsmath,amssymb,stmaryrd}
\usepackage{graphicx}
\usepackage{subfigure}  
\usepackage{hyperref}
\usepackage{epsfig}
\usepackage{mathrsfs}
\usepackage{verbatim}
\usepackage{centernot}

\newcommand{\ket}[1]{|#1\rangle}

\usepackage{array}

\usepackage{bm}	

\begin{document}

\title{Quantized transport induced by topology transfer between coupled one-dimensional lattice systems}

\author{Lukas Wawer}
\author{Rui Li}
\author{Michael Fleischhauer}
\affiliation{Department of Physics and Research Center OPTIMAS, University of Kaiserslautern, Germany}

\begin{abstract}
We show that a topological pump in a one-dimensional  (1D) insulator can induce a strictly quantized transport 
in an auxiliary chain of non-interacting fermions weakly coupled to the first.
The transported charge 
is determined by an integer topological invariant of
the ficticious  Hamiltonian of the insulator, given by the covariance matrix of single-particle correlations.
If the original system consists of non-interacting fermions, this number is
identical to the TKNN (Thouless, Kohmoto, Nightinghale, den Nijs)  invariant of the original
system and thus the coupling induces a transfer of topology to the auxiliary chain. When extended to particles with interactions, for which the TKNN number
does not exist, the transported charge in the auxiliary chain defines a  topological invariant for the interacting system.
In certain cases this invariant agrees with the many-body generalization of the TKNN number introduced by Niu, Thouless, and Wu (NTW). 
We illustrate the topology transfer  to the auxiliary system for
the Rice-Mele model of non-interacting fermions at half filling and the extended superlattice Bose-Hubbard model at quarter filling.
In the latter case the induced  charge pump is fractional.
\end{abstract}


\date{\today}
\maketitle

\section{introduction}

Topological states of matter \cite{Xiao-RMP-2010,Hazan-Kane-RMP-2010,Wen-RMP-2017} have fascinated physicists for a long time as they can give rise to interesting phenomena such as
protected edge states and edge currents, quantized bulk transport in insulating states and exotic elementary excitations 
\cite{Klitzing-PRL-1980,TKNN-PRL-1982,Tsui-PRL-1982,Laughlin-PRL-1983,Arovas-PRL-1984}. 
Free fermion systems with topological band structure are very well understood by now and a full classification
can be given in terms of generalized symmetries of the single-particle Hamiltonian \cite{Altland-PRB-1997,Schnyder-PRB-2008,Ryu-NJPhys-2010}.
An important sub-class of these systems are Chern insulators, such as the Harper-Hofstadter model, where time-reversal symmetry is
broken. A hallmark feature of them in the case of two spatial dimensions is the quantized Hall conductivity. A related phenomenon in one-dimensional systems 
is the quantized bulk transport upon adiabatic cyclic variations of system parameters \cite{Thouless-PRB-1983}. Both are strictly quantized in the thermodynamic limit and at zero 
temperature and are related to a topological invariant. 
In the last decade there has been a growing interest in the field of interacting topological insulators and a number of topological states have been discovered
that exist only because of interactions. Prime examples are  fractional quantum Hall states 
\cite{Laughlin-PRL-1983,Arovas-PRL-1984}
and their generalizations or the Haldane state in antiferromagnetic spin-1 
chains \cite{Haldane-PRL-1983}. While being incompressible like their non-interacting counterparts, interacting topological insulators are fundamentally different and are characterized by 
degeneracies and fractional topological charges. One of the challenges in the field of interacting topological insulators is to find suitable and measurable 
invariants to distinguish between topological phases. 

In the present paper we argue  that the covariance matrix of single-particle correlations can  be used to define directly measurable topological invariants of one-dimensional lattice models, with and without interactions and also at finite temperatures. 
In the absence of interactions all equilibrium
properties, including topological ones, are encoded in single-particle correlations. Interestingly, as shown in \cite{Bardyn-PRX-2018}, the latter extends also to finite temperature states. We here
show that topological properties encoded in the covariance matrix of an (interacting) one-dimensional lattice system can be transferred to a second auxiliary chain of non-interacting fermions giving rise to a quantized transport in the latter upon cyclic adiabatic variations of parameters. In mean-field approximation the dynamics of the auxiliary fermions is goverend by a single-particle Hamiltonian matrix, which is identical to the covariance matrix of the original model, which is therefore also termed
ficticious Hamiltonian.

If the 1D system are non-interacting fermions the transported charge in the auxiliary chain is just given by the TKNN (Thouless, Kohmoto, Nightinghale and den Nijs) invariant of the original fermion Hamiltonian \cite{TKNN-PRL-1982}.
If the original system is interacting, the number of auxiliary particles transported in a full cylce is still quantized and defines a topological invariant. 
We show that this invariant is fully determined by the Zak phase of the single-particle Bloch states of the ficticious Hamiltonian matrix. In many cases it is identical to the many-body generalization of the TKNN invariant by Niu, Thouless and Wu (NTW number) \cite{Niu-PRB-1985,Niu-JPhysA-1984}, but a general relation cannot be derived.

The idea to relate topological properties of interacting systems to single-particle quantities is not new. Building on early work by Volovik et al. \cite{Volovik-JETP-1988},
Wang, Qi, and Zhang \cite{Wang-PRL-2010,Wang-PRX-2012,Wang-PRB-2012} and Gurarie and coworkers \cite{Gurarie-PRB-2011,Manmanna-PRB-2012} suggested simplified topological invariants of interacting systems in terms of single-particle Green's functions. As is true for the ficticious Hamiltonian discussed in the present paper, the 
Green's function at zero frequency is simply related to the single-particle Hamiltonian in the case of non-interacting fermions. Also all generalized symmetries of the Hamiltonian are inherited by the Green's function in this case.

Our paper is organized as follows. In Sec. II we discuss the topological equivalence between the single-particle Hamiltonian of free fermions and the
covariance matrix in a gapped many-body ground state or in a finite-temperature state. In Sec. III we show that the topological properties of an auxiliary chain of fermions weakly coupled to the original one are  controlled by an effective single-particle Hamiltonian, the ficticious Hamiltonian above. Thus 
topological effects, such as quantized bulk transport, can be induced in the auxiliary chain. We illustrate our findings for a simple one-dimensional (1D) topological model, the Rice-Mele model (RMM) at half filling. The coupling scheme proposed in Sec. III is diagonal in momentum space and  difficult to implement. Therefore we discuss in Sec. IV an approximate implementation using  quasi-local couplings only. In Sec. V we extend the discussion to insulating ground states of
interacting  systems in 1D, including the case of degeneracy. As a specific example we consider the  superlattice Bose-Hubbard model with strong nearest and next-nearest neighbor interactions of bosons. This model has two degenerate, Mott-insulating ground states at quarter filling associated with a fractional topological charge.
Finally making use of the simple relation between the covariance matrices of a non-interacting fermion system at finite and  zero temperature, we give an outlook to 
measurements of finite-temperature topological invariants  \cite{Bardyn-PRX-2018,Bardyn-NJP-2013,Linzner-PRB-2016}.

\section{Free fermions: Topological equivalence of Hamiltonian and equilibrium covariance matrix}

\subsection{Model and equilibrium covariance matrix}

Let us first consider gapped ground states of non-interacting  fermions on a lattice with particle number conservation. 
We consider a one-dimensional lattice with lattice constant $a=1$ and $L$ unit cells and set $\hbar =1$ throughout this work. 
The operators $\hat c_{\mu,j},\hat c_{\mu,j}^\dagger$
describe the annihilation and creation of a fermion in the $j$th unit cell and with the index $\mu\in \{1,\dots,p\}$ denoting a possible internal degree of freedom within a unit cell. Assuming translational invariance for simplicity, the Hamiltonian can be written in second quantization in the form
\begin{equation}
H_{s}= \sum_k \sum_{\mu,\nu=1}^p  \hat c_\mu^\dagger(k) \, {\sf h}_{\mu\nu}(k) \,  \hat c_\nu(k).\label{eq:H}
\end{equation}
Some remarks about disordered systems will be made later. We consider a grand canonical setting with chemical potential $\mu$ which fixes the
total particle number in the system. We assume that $H_{s}$ has multiple bands and consider an insulator, i.e assume that $\mu$ is chosen within a
band gap of $H_{s}$. Topological properties of lattice models are characterized by integer-valued invariants, which in many cases are based on geometric phases,
such as the Zak phase \cite{Xiao-RMP-2010}, which is the  phase picked up by a Bloch state $\vert u_n(k)\rangle$ in a specific band $n$ upon parallel transport through the Brillouin zone. In the thermodynamic limit, where the lattice momentum becomes continuous it reads:
\begin{equation}
\phi_\textrm{Zak}^{(n)} = i \int_\textrm{BZ} dk \, \bigl\langle u_n(k)\bigr\vert \partial_k u_n(k)\bigr\rangle.
\end{equation}
The Zak phase is in general not unique as it depends on the gauge choice of the Bloch states $\vert u_n(k)\rangle$.
One-dimensional lattice models can possess non-trivial topological properties only when protected by symmetries.
A  well-known example is the Su-Shrieffer-Heeger(SSH) model \cite{SSH-PRL-1979}, where chiral symmetry 
enforces topology. The latter is characterized by a quantized value of the  Zak phase of $0$ or $\pi$ \cite{Xiao-RMP-2010}, which allows one to distinguish between two topologically different phases. These phase can only be connected by going through a phase-transition or breaking the symmetry. 
As suggested by Rice and Mele \cite{Rice-PRL-1982} adding an appropriate symmetry breaking term allows one to smoothly connect the phases avoiding the critical point. Adiabatic cyclic variations of the Hamiltonian parameters enclosing the critical point lead to a nontrivial winding of the Zak phase, which defines a topological invariant as integral over the two-dimensional parameter space $(k,t)$ on a torus
\begin{equation}
\nu_s = \frac{i}{2\pi}\int_0^\tau \!\!\! dt \int_\textrm{BZ} dk\, \bigl\langle \partial_t u_n(k)\bigl\vert \partial_k u_n(k)\bigr\rangle.\label{eq:winding}
\end{equation}
In two-dimensional systems with lattice momenta $k_x$ and $k_y$, one of the two momenta can take over the role of $t$. 

We now argue that the topological properties of the systems, determined by the single-particle Hamiltonian ${\sf h}(k)$ are also encoded in the 
covariance matrix of normal-ordered, single-particle correlations in a gapped many-body ground state:
\begin{equation}
{\sf m}_{\mu\nu}(k) =\langle  \hat c_\mu^\dagger(k)  \hat c_\nu(k)\rangle. 
\end{equation}
The  ground state of a system of free fermions on a lattice
is  a Gaussian state 
\begin{equation}
\rho = \frac{1}{Z} \exp\left\{-\sum_k \hat{\mathbf{c}}^\dagger(k)\, {\sf g}(k)\, \hat{\mathbf{c}}(k)\right\}\label{eq:Gauss}
\end{equation}
which is fully determined by a 
$p\times p$  Hermitian matrix ${\sf g}(k)$, and we have used the abbreviation $ \hat{\mathbf{c}}(k)=\bigl(\hat c_1(k),\dots, \hat c_p(k)\bigr)$.
 The  covariance matrix  of such a state is directly related to ${\sf g}$  by \cite{Bardyn-NJP-2013}
\begin{equation}
{\sf m}_{\mu\nu}(k) = \frac{1}{2}\left[ 1- \tanh\left(\frac{{\sf g}(k)}{2}\right)\right]_{\mu,\nu}.\label{eq:m}
\end{equation}

One notices that eq.(\ref{eq:Gauss}) has the form of a Gibbs state and indeed an 
equilibrium state of $H_{s}$ at finite $\beta=1/(k_BT)$ is 
also a Gaussian state with
\begin{equation}
{\sf g}(k) = \beta \Bigl( {\sf h}(k) -\mu\Bigr).\label{eq:g}
\end{equation}
The ground state is obtained in the limit $\beta \to \infty$. Most importantly all (single-particle) eigenstates $\vert \epsilon_n(k)\rangle$ of ${\sf h}(k)$ are also eigenstates of 
${\sf g}(k)$ and thus of the covariance matrix
${\sf m}(k)$.

\subsection{Ficticious Hamiltonian}
\label{sec:fict-H}

Let us now consider a free-fermion lattice system with ficticious Hamiltonian \cite{Bardyn-NJP-2013}
\begin{eqnarray}
H_\textrm{fict} = \eta \sum_k \sum_{\mu,\nu=1}^p  \hat a_\mu^\dagger(k) \, {\sf m}_{\mu\nu}(k) \,  \hat a_\nu(k)\label{eq:H-fict}
\end{eqnarray}
where $\hat a_\mu(k)$ and $\hat a_\mu^\dagger(k)$ are fermion annihilation and creation operators in momentum space and 
${\sf m}(k)$ is the single-particle covariance matrix, eq.\eqref{eq:m}. 

If  ${\sf m}(k)$ corresponds to a gapped ground state
of a fermion lattice model, the ficticious  Hamiltonian ${\sf h}_\textrm{fict}(k) = \eta {\sf m}(k)$
is also gapped. In an insulating non-degenerate ground state of the original system, 
the ficticious Hamiltonian ${\sf h}_\textrm{fict}(k)$  has a flat spectrum
\begin{equation}
\epsilon_n^\textrm{fict}(k) =\Biggl\{\begin{array}{ll} 0 & \, \textrm{if}\,\, \epsilon_n(k) > \mu\\
\eta &\, \textrm{if}\,\,  \epsilon_n(k) < \mu
\end{array}.
\end{equation}
with only two energy "bands". 
Note that for positive values of $\eta$ the spectrum of the ficticious single-particle Hamiltonian is reversed with respect to that of
${\sf h}(k)$. 
Depending on the sign of $\eta$, the gapped many-body ground state of the ficticious Hamiltonian thus  contains  all single-particle eigenstates with energies $\epsilon_n(k)$ either above (for $\eta >0$) or below (for $\eta <0$) the chemical potential $\mu$. In particular if $\eta>0$ and the 
original Hamiltonian has $d$ bands with energy $\epsilon_n(k)<\mu$, ${\sf h}_\textrm{fict}(k)$ has a $p-d$-fold degenerate ground state for every lattice momentum $k$ with energy $\epsilon^\textrm{fict}_n(k)=0$ and a $d$-fold degenerate excited state with energy $\epsilon^\textrm{fict}_n(k)=\eta$.
If $\eta <0$ the situation is reversed.

We now argue that the ficticious system inherits the topological properties of the original one. Let us consider
adiabatic parameter variations of the system Hamiltonian $H_{s}(t)$ in a closed loop in time from $t=0$ to $t=\tau$, such that 
$H_{s}(t)=H_{s}(t+\tau)$. Then the initial and final states of the original
system are the same apart from a phase and also the ground state of the ficticious system will return to itself.
Since the single-particle eigenstates of the ficticious Hamiltonian are identical to those of the system Hamiltonian,
 the  Wilson loop \cite{Wilczek-PRL-1984}
 of the $d$-fold ($\eta <0$)  degenerate ground-state manifold of $H_\textrm{fict}$
\begin{eqnarray}
\nu_\textrm{fict} &=& \frac{1}{2\pi}\int_0^\tau \!\!\!\! dt\, \frac{\partial}{\partial t} \textrm{Im} \log \det {\sf W}  \nonumber\\
&=& 
 \frac{1}{2\pi}\int_0^\tau \!\!\!\! dt\, \frac{\partial}{\partial t} \textrm{Im}\,  \textrm{Tr} \log {\sf W},\quad
\end{eqnarray}
with the $d\times d$  matrix $(n,m\in \{1,\dots,d\})$
\begin{equation}
{\sf W}_{nm} = \exp\biggl\{i\int_\textrm{BZ}\!\! dk\, \bigl\langle u_n(k)\bigl\vert \partial_k u_m(k)\bigr\rangle\biggr\},
\end{equation}
which is identical to the winding of the total Zak phase
of all occupied bands (for $\eta <0$)
\begin{equation}
\nu_s = \frac{i}{2\pi}\int_0^\tau \!\!\! dt\sum_{n; \textrm{occup.}}\int_\textrm{BZ} dk\, \bigl\langle \partial_t u_n(k)\bigl\vert \partial_k u_n(k)\bigr\rangle.\label{eq:Zak-system}
\end{equation}
Here the  $\vert u_n(k)\rangle$ are the Bloch states of the original 
Hamiltonian (\ref{eq:H}), with $\langle \mathbf{r}\vert \epsilon_n(k) \rangle = e^{i\mathbf{k}\cdot\mathbf{r}} \langle \mathbf{r}\vert u_n(k)\rangle$,
which are also eigenstates of the ficticious Hamiltonian  (\ref{eq:H-fict}).
For $\eta>0$ $\nu_\textrm{fict}$ corresponds to the winding number of all unoccupied bands $\tilde \nu_s = -\nu_s$
of the original system. 
That the two Hamiltonians $H$ and $H_\textrm{fict}$ have the same topological 
properties in their corresponding insulating ground states is no surprise as one recognizes from eqs.(\ref{eq:m}) and (\ref{eq:g}) 
that ${\sf h}(k)$ can be smoothly deformed into ${\sf m}(k)$ without closing the many-body gap and thus both are topologically equivalent.

\section{Topology transfer from a free-fermion system to an auxiliary system}
\label{free}

We now want to show that the ficticous Hamiltonian of a 1D system can be physically realized by a  weak coupling to
an auxiliary chain of otherwise non-interacting fermions. In this way topological properties are transferred from one system
to a second, auxiliary one. This "topology transfer" manifests itself e.g. in a quantized charge transport
in the  auxiliary system upon periodic adiabatic modulations of the  original Hamiltonian. 

In this section we consider as the ''system'' non-interacting fermions in a gapped many-body ground state, which
is non-degenerate.
The generalization to interacting fermions, which also includes the possibility of degeneracies, will be discussed in a subsequent section.

\begin{figure}[htb]
	\begin{center}
	\includegraphics[width=0.8\columnwidth]{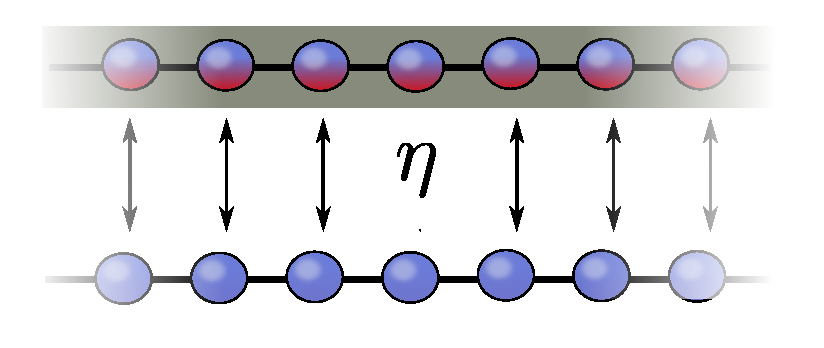}
	\end{center}
	\caption{Sketch of topology transfer scheme: The original  chain of free or interacting fermions (top) is weakly coupled to an auxiliary system of non-interacting fermions (bottom). The coupling is diagonal in momentum space and conserves the particle numbers in both chains.}
	\label{fig:transfer}
\end{figure}

\subsection{Model}

To be specific we consider two one-dimensional chains of fermions weakly coupled to each other as indicated in Fig.\ref{fig:transfer}.
The system, represented by the top chain, is described by a free-fermion Hamiltonian  $H_{s}$ with annihilation and creation operators $\hat c_\mu(k)$ and $\hat c^\dagger_\mu(k)$, where $k$ is the lattice momentum and 
$\mu$ labels the degrees of freedom within a unit cell. It is weakly coupled to an ''auxiliary'' system of otherwise non-interacting fermions 
with respective annihilation and creation operators $\hat a_\mu(k)$ and $\hat a^\dagger_\mu(k)$ according to
\begin{eqnarray}
H &=& H_{s} + H_\eta ,\\
H_\eta &=& \eta \sum_k \sum_{\mu,\nu=1}^p   \hat  c_\mu^\dagger(k) \hat c_\nu(k) \hat a_\mu^\dagger(k) \hat a_\nu(k).\label{eq:H_eta}
\end{eqnarray}
Here we have assumed a unit cell of $p$ sites.
The number of fermions in both chains is individually conserved and we assume that it is chosen such that the combined system has an insulating many-body ground state $\vert \Phi_0\rangle$. If $\vert\eta\vert$ is small compared to the gap of $H_{s}$, then the original system is only little affected by the coupling and is approximately described by its
ground state $\vert \Phi_0^{s}\rangle$, which is also insulating. 

 For the following discussion it useful to rewrite the total Hamiltonian in the form
\begin{equation}
H=H_0 + H_1 \label{eq:H-alt}
\end{equation}
where
\begin{equation}
H_0 = H_{s} + \eta \sum_k \sum_{\mu,\nu=1}^p   \bigl\langle \hat  c_\mu^\dagger(k) \hat c_\nu(k)\bigr\rangle  \hat a_\mu^\dagger(k) \hat a_\nu(k)\label{eq:H0}
\end{equation}
contains the system Hamiltonian and the mean-field interaction Hamiltonian, where ${\sf m}_{\mu\nu}= \langle \hat  c_\mu^\dagger(k) \hat c_\nu(k)\rangle$ is evaluated
in the ground state $\vert \Phi_0^{s}\rangle$ of $H_{s}$. Thus eq.\eqref{eq:H0} represents the ficticious Hamiltonian (\ref{eq:H-fict}) experienced by the auxiliary fermions. The second term in (\ref{eq:H-alt})
formally describes the coupling of the auxiliary system to fluctuations in the original system
\begin{equation}
H_1 =  \eta \sum_k \sum_{\mu,\nu=1}^p   \Bigl(\hat  c_\mu^\dagger(k) \hat c_\nu(k) - \bigl\langle \hat  c_\mu^\dagger(k) \hat c_\nu(k)\bigr\rangle\Bigr)
\hat a_\mu^\dagger(k) \hat a_\nu(k)
\end{equation}
and is responsible for the buildup of entanglement between the two fermion chains.

\subsection{Topology transfer and induced quantized particle transport}
\label{sect-B}

We now argue that an adiabatic cyclic modulation of $H_{s}\to H_{s}(t)$ in time will lead to
a topological charge pump in the auxiliary system. 
A modulation $H_s \to H_s(t)= H_s(t+\tau)$ in time with period $\tau$ 
implies a corresponding cyclic modulation of the covariance matrix ${\sf m}(t)$ and the ficticious Hamiltonian. 
As we will show this gives rise to an adiabatic  transport of $Q_a$ particles in the auxiliary system, and $Q_a$ is equal to the 
total winding number of occupied bands of the original system, $\nu_s$, for an attractive coupling $\eta<0$, or the total winding number
of unoccupied bands, $\tilde \nu_s=-\nu_s$, for the case of a repulsive coupling $\eta >0$:
\begin{equation}
Q_a =\left\{ \begin{array}{c} 
+\nu_s,\quad\textrm{for}\, \eta <0\\
-\nu_s,\quad\textrm{for}\, \eta >0
\end{array}\right.
\end{equation}

To calculate the number of transported particles $Q_a$ in the auxiliary system, we follow the procedure of Niu, Thouless and Wu \cite{Niu-PRB-1985}
and integrate the expectation value of the total momentum operator of the
auxiliary fermions $\hat P_a = \sum_{i=1}^N \hat p_i$ in a non-degenerate adiabatic eigenstate $\vert \Phi(t)\rangle$ with $\vert \Phi(t=0)\rangle = \vert \Phi_0\rangle$ of the total system. Since there is no transport in instantaneous insulating eigenstates, one has to consider the lowest-order non-adiabatic correction
\begin{eqnarray}
\vert \Phi(t)\rangle  = \vert \Phi_0(t)\rangle+ 
 i \sum_{n\ne 0} \frac{\vert \Phi_n(t)\rangle \langle \Phi_n(t)\vert \partial_t \Phi_0(t)\rangle}{E_n(t)-E_0(t)}.\qquad 
\end{eqnarray}
$\vert \Phi_n(t)\rangle$ are the  excited eigenstates with energy $E_n(t)$.
This gives in lowest order of non-adiabatic corrections
\begin{eqnarray}
Q_a &=& \frac{1}{L} \int_0^\tau\!\! dt\, \langle \Phi(t)\vert \hat P_a \vert \Phi(t)\rangle,\label{eq:Q}\\
&=&  \frac{1}{L} \int_0^\tau\!\! dt\, \sum_{n\ne 0} 
\frac{\langle \Phi_0(t)\vert \hat P_a \vert \Phi_n(t)\rangle \langle \Phi_n(t)\vert \partial_t \Phi_0(t)\rangle}{E_n(t)-E_0(t)} +c.c..\nonumber
\end{eqnarray}
In order to calculate the matrix elements of the momentum operator  it is useful to perform a canonical transformation of the Hamiltonian $H(\alpha) = e^{-i\alpha \hat X_a} H e^{i\alpha \hat X_a}$, with $\hat X_a$ being the position operator of the particles of the auxiliary system. Making use of $\hat P_a = - i [\hat X_a, H] = \partial_\alpha H(\alpha)\bigr\vert_{\alpha=0}$ one finds 
\begin{eqnarray}
&& \langle \Phi_0 \vert \hat P_a\vert \Phi_n\rangle = \langle \Phi_0 \vert \frac{\partial H(\alpha)}{\partial \alpha}  \vert \Phi_n\rangle\biggr\vert_{\alpha =0} \nonumber\\
&&\quad =
- \langle \partial_\alpha \Phi_0(\alpha) \vert \Phi_n(\alpha)\rangle \Bigl(E_n(\alpha)-E_0(\alpha)\Bigr)\biggr\vert_{\alpha =0}.
 \end{eqnarray}
This gives for the transported charge 
\begin{equation}
Q_a = -\frac{i}{L} \int_0^\tau\!\! dt\, \Bigl(\bigl\langle \partial_\alpha \Phi_0\bigl\vert \partial_t \Phi_0\bigr\rangle - 
\bigl\langle \partial_t \Phi_0 \bigl\vert \partial_\alpha \Phi_0\bigr\rangle\Bigr)_{\alpha =0}.
\end{equation}
Following Ref.\cite{Niu-PRB-1985}
 $Q_a$ is identical to its average over all values of $\alpha$ between $\{-\pi /L,\pi/L\}$ in the thermodynamic limit
$L\to \infty$. Note that $\alpha =\pm \pi/L$ correspond to the same situation.
This  gives an integral  of a Berry curvature over a two-dimensional torus 
\begin{equation}
Q_a= -\frac{i}{2\pi} \int_0^\tau\!\! dt\, \int_{-\pi/L}^{\pi/L}\!\!\! d\alpha \Bigl(\bigl\langle \partial_\alpha \Phi_0 \bigl\vert \partial_t \Phi_0\bigr\rangle - 
\bigl\langle \partial_t \Phi_0 \bigl\vert \partial_\alpha \Phi_0\bigr\rangle\Bigr) \label{eq:transport}
\end{equation}
which must be an integer and is a topological invariant $\nu_a$, corresponding to a winding number.

We now show that this winding number is identical to the topological invariant characterizing all occupied bands (respectively all unoccupied bands) of the
original system.  To this end we assume  a weak coupling $\eta$ where $H_1$ can be considered as a perturbation to $H_0$. For simplicity we
assume in the rest of this subsection $\eta < 0$. All arguments can however straightforwardly be applied to $\eta >0$.

\subsubsection{Zeroth order perturbation in $H_1$}

 In lowest order of $H_1$ the evolution of the system chain is unaffected by the auxiliary one and 
 the dynamics of the latter is determined by the mean-field (ficticious) Hamiltonian 
 \begin{equation}
 H_\textrm{fict}(t) = \eta \sum_k \sum_{\mu,\nu=1}^p  \hat a_\mu^\dagger(k) \, {\sf m}_{\mu\nu}(k,t) \,  \hat a_\nu(k).\label{eq:H-fict2}
 \end{equation}
 Since the Hamiltonian of the system chain ${\sf h}_{\mu\nu}$ is modulated in time with period $\tau$, the
 covariance matrix is also $\tau$-periodic, ${\sf m}_{\mu\nu}(k,t+\tau) ={\sf m}_{\mu\nu}(k,t)$. 
In lowest order perturbation there is no buildup of entanglement between the two subsystems
and all eigenstates of the combined system factorize. In particular one has for the ground state of the total system
\begin{equation}
\vert \Phi_0^{(0)}\rangle = \vert \Phi_0^s \rangle \, \vert \Phi_0^a\rangle
\end{equation}
where $\vert \Phi_0^a\rangle$ is the ground state of (\ref{eq:H-fict2}). The transported charge $Q_a$ is thus given by eq.\eqref{eq:transport},
with $\vert \Phi_0\rangle$ replaced by the ground state $\vert \Phi_0^a\rangle$ of $H_\textrm{fict}$.
 Since for non-interacting fermions the covariance matrix has the same eigenfunctions as the single-particle Hamiltonian ${\sf h}_{\mu\nu}(k,t)$ of the original system, 
 $Q_a$  is given in lowest order perturbation by the winding number of the total Zak phase  of
 the system Hamiltonian, i.e. 
 \begin{equation}
Q_a^{(0)} = \nu_\textrm{fict} = \left\{ \begin{array}{c} 
+\nu_s,\quad\textrm{for}\, \eta <0\\
-\nu_s,\quad\textrm{for}\, \eta >0
\end{array}\right.
 \end{equation}
 The excited states with energy $E_{n,m}^{(0)}$ can be labelled with two indices $n$ and $m$ corresponding to the 
system and auxiliary chain respectively
\begin{equation}
\vert \Phi_{n,m}^{(0)}\rangle = \vert \Phi_n^s \rangle \, \vert \Phi_m^a\rangle.
\end{equation}
%
 
\subsubsection{First order perturbation in $H_1$}

 In first order of $H_1$ the instantaneous ground state of the combined system reads
 \begin{eqnarray}
\vert \Phi_0^{(1)}\rangle  = \sqrt{p} \vert \Phi_0^{(0)}\rangle+ \sqrt{1-p} \vert \tilde \Phi\rangle\label{eq:Psi1}
\end{eqnarray}
where $\sqrt{p}$ is the overlap between the exact ground state of the system and the unperturbed one. The normalized
correction to the state vector reads
\begin{equation}
 \vert \tilde \Phi\rangle = \sqrt{\frac{p}{1-p}} 
 \sum_{n\ne 0}\sum_m \frac{\vert \Phi_{n,m}^{(0)}\rangle \langle \Phi_{n,m}^{(0)} \vert H_1 \vert \Phi_0^{(0)}\rangle}{E_{n,m}^{(0)}-E_0^{(0)}}
 \end{equation}
 Since $H_1 \sim  \bigl(\hat c_\mu^\dagger \hat c_\nu - \langle \Phi_0^s\vert \hat c_\mu^\dagger \hat c_\nu \vert \Phi_0^s\rangle\bigr)$,
the only  states contributing to $\vert \tilde \Phi\rangle$ are those where the
system chain is excited  $\vert \Phi_{n>0}^s\rangle \vert \Phi_m^a\rangle$.
Thus the denominator is always larger than the energy gap of the system, i.e. $E_{n,m}^{(0)} - E_0^{(0)} \ge \Delta_\textrm{gap}$, and
the probability $1-p$ for the exact ground state to contain components orthogonal to the unperturbed one scales as
\begin{eqnarray}
(1-p) &\sim& \sum_{n\ne 0}\sum_m \frac{\langle \phi_0^{(0)}\vert H_1\vert \Phi_{n,m}^{(0)}\rangle \langle \Phi_{n,m}^{(0)} \vert H_1 \vert \Phi_0^{(0)}\rangle}{(E_{n,m}^{(0)}-E_0^{(0)})^2}\nonumber\\
&\sim &{\cal O}\left(
\frac{\eta^2}{\Delta^2_\textrm{gap}}\right).
\end{eqnarray}
Plugging $\vert \Phi_0^{(1)}\rangle$ into  expression \eqref{eq:transport} for the transported charge yields
\begin{eqnarray}
 &&Q_a^{(1)}= -\frac{i}{2\pi} \int\! dt\int \! d\alpha\, \biggl\{\bigl\langle \partial_\alpha \Phi_0^{(1)} \bigl\vert \partial_t \Phi_0^{(1)} \bigr\rangle -c.c.\biggr\}\label{eq:Q1}\\
&&\qquad = -\frac{i}{2\pi} \int\! dt\int \! d\alpha\,  \biggl\{p \Bigl(\bigl\langle \partial_\alpha \Phi_0^{(0)} \bigl\vert \partial_t \Phi_0^{(0)} \bigr\rangle -c.c.\Bigr)+ \nonumber\\
 &&\qquad\qquad\qquad\qquad +(1-p) \Bigl( \langle\partial_\alpha\tilde \Phi\vert \partial_t \tilde \Phi \rangle -c.c.\Bigr)+\label{eq:Q1b}\\
 &&\quad +\sqrt{p(1-p)}\Bigl(\bigl\langle \partial_\alpha \Phi_0^{(0)} \bigl\vert \partial_t \tilde\Phi \bigr\rangle
 +\bigl\langle \partial_\alpha \tilde\Phi \bigl\vert \partial_t \Phi_0^{(0)} \bigr\rangle - \nonumber\\
 &&\qquad\qquad\qquad\quad - \bigl\langle \partial_t \tilde\Phi \bigl\vert \partial_\alpha \Phi_0^{(0)} \bigr\rangle
 -\bigl\langle \partial_t \Phi_0^{(0)} \bigl\vert \partial_\alpha \tilde\Phi \bigr\rangle
 \Bigr)\biggr\}.\nonumber
\end{eqnarray}
As can be seen from eq.\eqref{eq:Q1}, also
$Q_a^{(1)}$ is an integral of a Berry curvature of a gapped many-body state over a two-dimensional torus and thus an integer.
Apart from the prefactors $p$ and $1-p$ the same holds for the integrals in the 
second and third line of the above expression. 
Furthermore, as we will show in the Appendix,  the last integral in \eqref{eq:Q1b} vanishes exactly. 
%
%
%
%
Thus we can write
\begin{equation}
Q_a^{(1)} =  Q_a^{(0)} + (1-p) \tilde Q_a,\label{eq:Qa1}
\end{equation}
%
with $\tilde Q_a$ being an integer.
As long as $\eta$ is sufficiently small compared to the gap of the unperturbed system, $\Delta_\textrm{gap}$, there is no phase transition in the
combined system and thus the winding number of the total system, as well as the transported charge $Q_a^{(1)}$ must be the same as in the
limit $(1-p)\to 0$. The only integer in eq.\eqref{eq:Qa1} compatible with this is $\tilde Q_a\equiv 0$.
Thus also in first order perturbation one has
\begin{equation}
Q_a^{(1)} = \nu_\textrm{fict} = \left\{ 
\begin{array}{c} 
+\nu_s,\quad\textrm{for}\, \eta <0\\
-\nu_s,\quad\textrm{for}\, \eta >0
\end{array}\right. ,
\end{equation}
i.e. a cyclic adiabatic variation of parameters of the orginial system Hamiltonian induces a strictly quantized topological charge transport in the auxiliary system.

\subsection{Example}

\begin{figure}[htb]
	\begin{center}
	\includegraphics[width=0.75\columnwidth]{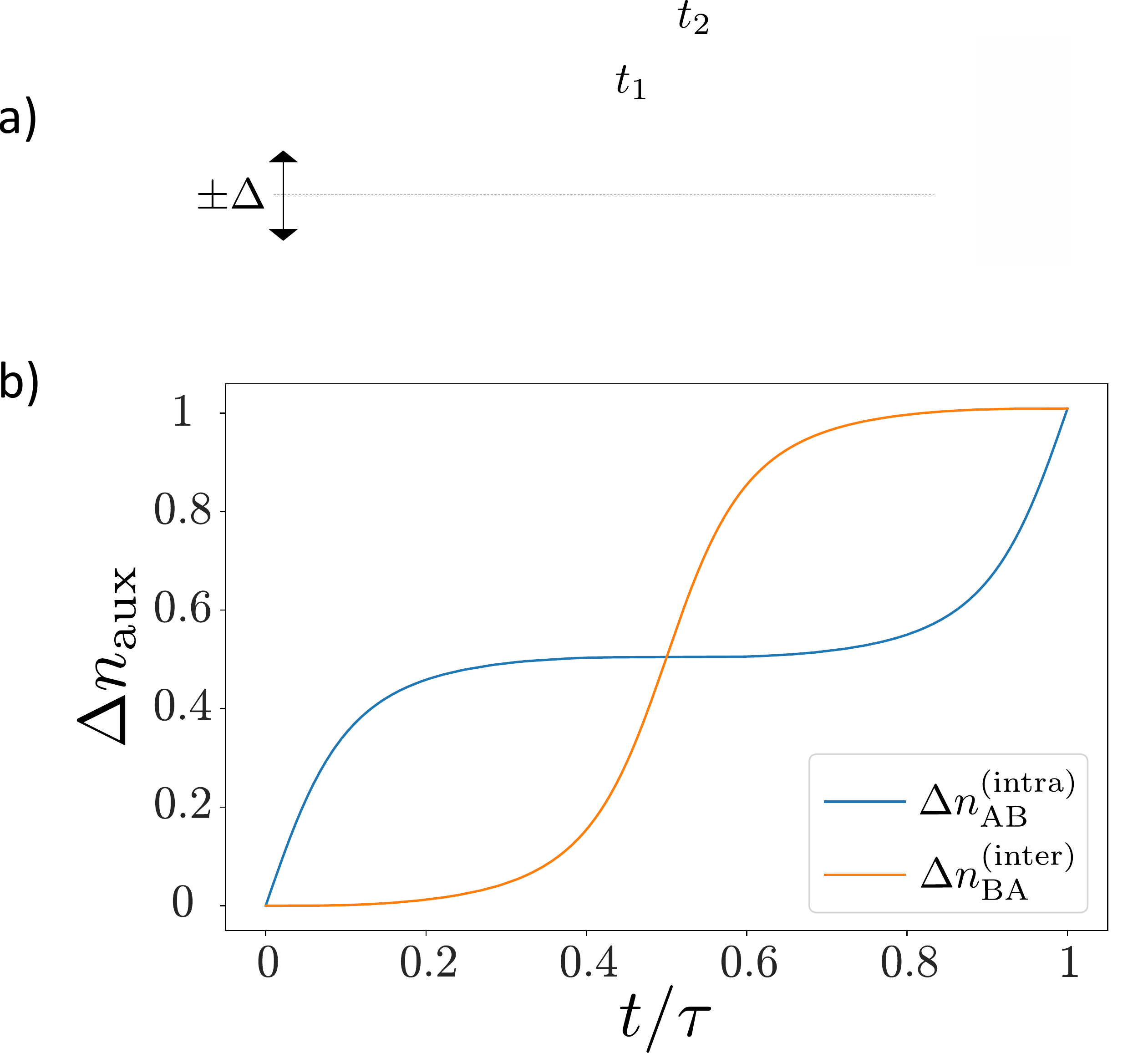}
	\end{center}
	\caption{(a) Rice-Mele model (RMM) with alternating hopping $t_1$, $t_2$ and staggered potential $\pm \Delta$. (b) Particle transport in auxiliary chain    
	between adjacent unit cells $(\Delta n_\textrm{BA}^\textrm{inter})$ and
	within unit cell $(\Delta n_\textrm{AB}^\textrm{intra})$ as function of time upon adiabatic change of parameters of RMM. The curves, which are plotted for $\eta/\Delta_\mathrm{gap}^\mathrm{RM}= 0.1; 1; 10$ lie exactly on top of each other. Here $\Delta_\mathrm{gap}^\mathrm{RM}$ is the minimum gap along the parameter path of the RMM:
	$\Delta(t) = -6 \sin(2\pi\frac{t}{\tau})$, $t_1(t)= 2(1+\cos(2\pi\frac{t}{\tau}))$, $t_2(t)=2(1-\cos(2\pi\frac{t}{\tau}))$.}
	\label{fig:system}
\end{figure}

To illustrate the topology transfer let us consider the simplest topologically non-trivial  model in 1D, the Rice-Mele model (RMM) \cite{Rice-PRL-1982}. 
This model, which is sketched in Fig. \ref{fig:system}a, describes lattice fermions in the tight-binding limit with alternating tunnel couplings
$t_1$ and $t_2$ between neighboring sites and a staggered on-site potential $\pm \Delta$. The unit cell thus contains two sites, labeled $A$ and $B$. 
The Hamiltonian can be written in momentum space as
\begin{align}
	& \quad H_\mathrm{RM} =\label{eq:RMM} \\
		&= \sum_k \begin{pmatrix}
		\hat{c}_{\mathrm{A}}^\dagger(k) \\
		\hat{c}_{\mathrm{B}}^\dagger(k)
		\end{pmatrix}^\mathrm{T}
		\begin{pmatrix}
		\Delta & -t_1-t_2 \mathrm{e}^{-ik} \\ 
		 -t_1-t_2 \mathrm{e}^{ik} & -\Delta
		\end{pmatrix}
		\begin{pmatrix}
		\hat{c}_{\mathrm{A}}(k)\\
		\hat{c}_{\mathrm{B}}(k)
		\end{pmatrix}.\nonumber
	\end{align}
Here $\hat c_\mathrm{A,B}(k)$ denote the fermion annihilation operators of the $\mathrm{A}$ or $\mathrm{B}$ sublattice in momentum space. The RMM has two bands 
\begin{equation}
		\varepsilon_\pm(k) = \pm \varepsilon(k) = \pm \sqrt{\Delta^2 +t_1^2+t_2^2+2t_1t_2\cos(k)}
\end{equation}
and the gap
closes only for $\Delta=t_1-t_2=0$. 
Except at this  point in parameter space, the RMM has 
thus an insulating ground state at half filling. Adiabatically changing $\Delta=\Delta(t)$
as well as $t_{1,2}=t_{1,2}(t)$ in a closed loop encircling the degeneracy point leads to a non-trivial winding of the Zak phase of the lower (upper) band, 
$\nu_s=+1$ ($\tilde \nu_s =-1$),
associated with
a quantized topological transport of one particle.

We now couple the RMM to an auxiliary chain of otherwise non-interacting fermions at half filling, according to eq.(\ref{eq:H_eta}).
If we choose, say, a positive value of $\eta$, we expect a quantized transport of particles in the auxiliary chain 
in the opposite direction as in the Rice-Mele ground state.
Figure \ref{fig:system}b shows numerical simulations of the particle transport in the  auxiliary chain as function of time for different values of the coupling $\eta$ ($\eta/\Delta_\textrm{gap}^\textrm{RM} = 0.1, 1, 10$). As expected one recognizes  anti-parallel quantized transport. On first glance it is surprising that the transport in the auxiliary chain
remains strictly quantized to one also beyond the perturbative limit of small $\eta$. This can be understood as follows: The Hamiltonian of the total system, consisting of the RMM
and the auxiliary chain, can be written in the form
\begin{widetext}
	\begin{equation}
	{H} = \sum_k \begin{pmatrix}
	\hat{c}_{\mathrm{A}}^\dagger(k) \\
	\hat{c}_{\mathrm{B}}^\dagger(k)
	\end{pmatrix}^\mathrm{T}
	\begin{pmatrix}
	\Delta + \eta \hat{a}_{\mathrm{A}}^\dagger(k) \hat{a}_{\mathrm{A}}(k) & \left(-t_1-t_2 \mathrm{e}^{-ik}\right) + \eta \hat{a}_{\mathrm{A}}^\dagger(k) \hat{a}_{\mathrm{B}}(k) \\ 
	\left(-t_1-t_2 \mathrm{e}^{ik}\right) +\eta \hat{a}_{\mathrm{B}}^\dagger(k)\hat{a}_{\mathrm{A}}(k) & -\Delta + \eta \hat{a}_{\mathrm{B}}^\dagger(k) \hat{a}_{\mathrm{B}}
	\end{pmatrix}
	\begin{pmatrix}
	\hat{c}_{\mathrm{A}}(k)\\
	\hat{c}_{\mathrm{B}}(k)
	\end{pmatrix},
	\label{ferm_basis}
	\end{equation}
where $\hat a_\mathrm{A,B}(k)$ denote the fermion annihilation operators of the $\mathrm{A}$ or $\mathrm{B}$ sublattice in the auxiliary system. The total Hamiltonian conserves the
particle number in each subsystem and for every momentum. Thus  we can express \eqref{ferm_basis}  at double half filling in 
the particle-number basis $\ket{{n}_\mathrm{A}^c {n}_\mathrm{B}^c {n}_\mathrm{A}^a {n}_\mathrm{B}^a}_k$ as follows
	\begin{equation}
	H = \sum_k \begin{pmatrix}
	|1001\rangle \\
	|0101\rangle \\
	|1010\rangle \\
	|0110\rangle
	\end{pmatrix}^\mathrm{T}
	\begin{pmatrix}
	\Delta   & -\left(t_1+t_2 \mathrm{e}^{-ik}\right) & 0 & 0  \\ 
	-\left(t_1+t_2 \mathrm{e}^{ik}\right)  & -\Delta + \eta & \eta & 0 \\
	0 & \eta & \Delta + \eta & -\left(t_1+t_2 \mathrm{e}^{-ik}\right) \\
	0 & 0 & -\left(t_1+t_2 \mathrm{e}^{ik}\right) & -\Delta
	\end{pmatrix}
	\begin{pmatrix}
	\langle 1001| \\
	\langle 0101| \\
	\langle 1010| \\
	\langle 0110|
	\end{pmatrix}.
	\label{particle_basis}
	\end{equation}
\end{widetext}
The $4\times 4$ Hamiltonian has 4 eigenvalues for every lattice momentum, which can easily be calculated
	\begin{align}
	&\varepsilon_\pm(k) =\pm \varepsilon(k)=\pm \sqrt{\Delta^2+t_1^2+t_2^2+2t_1t_2\cos(k)}, \label{ev_RM} \\
	&\varepsilon_\pm(k,\eta) = \eta \pm \sqrt{\varepsilon(k)^2 +\eta^2} \label{ev_aux}.
	\end{align}
The corresponding bands are separated by gaps, which  close for any $\eta\ne 0$ only
for $\Delta=t_1-t_2=0$. There is no band crossing because for $\eta>0$:  $\varepsilon_+(k,\eta) > \varepsilon_+(k) > 0 >  \varepsilon_-(k,\eta) >\varepsilon_-(k)$.
In the absence of a gap-closing upon changing $\eta$, there is no topological phase transition and thus the transported charge in both systems remains
quantized to the same value for arbitrary values of $\eta$, i.e. in particular also beyond the perturbative regime.  
The many-body gap $\Delta_\mathrm{gap}$ of the full system at double half filling ranges between $\vert\eta\vert$ and half the gap of the RMM, $\Delta^\textrm{RM}_\mathrm{gap}$, with
	\begin{equation}
	\Delta_\mathrm{gap} \simeq \Biggl\{
	\begin{array}{ll} \vert\eta\vert & \textrm{for}\quad \vert\eta\vert \ll 1\\
	\frac{\Delta_\mathrm{gap}^\mathrm{RM}}{2}  & \textrm{for}\quad \vert\eta\vert \gg 1.
	\end{array}
	\end{equation}
Another surprising feature of the simulations in Fig.\ref{fig:system}b and c, is, that all curves lie on top of each other. This is due to the choice of $\eta$ being positive. In this case 
the lowest eigenstate of \eqref{particle_basis}, $\vert \phi_-\rangle$, does not depend on $\eta$: 
\begin{eqnarray}
|\phi_\pm\rangle &\sim&  -\frac{(\Delta\pm\varepsilon(k))^2}{(t_1+t_2e^{ik})^2}|1001\rangle+\frac{\Delta\pm\varepsilon(k)}{t_1+t_2e^{ik}}|0101\rangle\nonumber\\
&&  -\frac{\Delta\pm\varepsilon(k)}{t_1+t_2e^{ik}}|1010\rangle+|0110\rangle,\\
|\phi_\pm^\eta\rangle &\sim& \frac{t_1+t_2e^{-ik}}{t_1+t_2e^{ik}}|1001\rangle+\frac{\Delta-\varepsilon_\pm(k,\eta)}{t_1+t_2e^{ik}}|0101\rangle\nonumber\\
&& -\frac{\Delta+\varepsilon_\pm(k,\eta)}{t_1+t_2e^{ik}}|1010\rangle+|0110\rangle.
\end{eqnarray}
 Thus not only the net charge transported in one full cycle of the Thouless pump is exactly quantized, the transport is completely independent on $\eta$.
 
%

\subsection{Comment on the effect of disorder}

Until this point we have assumed lattice translational invariance, which excludes the presence of disorder. Topological properties are however expected to be robust against weak perturbations and thus we have to discuss if the
topology transfer survives in the presence of disorder. To this end we can modify the discussion in Sec. \ref{sect-B} and add a
disorder potential acting on the original system 
\begin{equation}
H_1 \, \to\,  H_1 + H_\mathrm{dis}.
\end{equation}
We can repeat the perturbation arguments given in Sec. \ref{sect-B} and see that as long as $H_\textrm{dis}$ does not lead to a gap closing,
the transported charge in the auxiliary system remains strictly quantized to the value set by the ficticious Hamiltonian.

\section{Implementation of topology transfer}

The  coupling Hamiltonian \eqref{eq:H_eta}, which induces the topology transfer between system and auxiliary chain  is diagonal in momentum space
and as such difficult to realize. In the following we will show that it can be approximately implemented by couplings that are short-ranged in
real space and thus can be  realized e.g. in experiments with ultra-cold gases. We will restrict the following discussion to a system with
translational invariance of the ground state by two lattice sites. Then transforming to a real-space description via
\begin{equation}
\hat c_\mu(k) = \frac{1}{\sqrt{L}} \sum_{j=1}^L e^{\frac{2\pi ij k}{L}} \hat c_{j,\mu}
\end{equation}
where the index $j\in\{1,\dots,L\}$ denotes the unit cell and $\mu\in\{\mathrm{A},\mathrm{B}\}$ the intra-cell degree of freedom, the ficticious Hamiltonian \eqref{eq:H-fict} can be expressed as
\begin{equation}
H_\textrm{fict}= \eta \sum_{m,n} \sum_{\mu,\nu} \Bigl(\langle \hat c^\dagger_{m,\mu} \hat c_{n,\nu}\rangle \hat a_{m,\mu}^\dagger
\hat a_{n,\nu} + h.a. \Bigr).
\end{equation}
Here we have used the translational invariance of  ground-state correlations. In an insulating  state of the original system, off-diagonal
first-order coherences decay exponentially  with distance and to good approximation it is  sufficient to consider  nearest neighbor correlations. 
If we denote the left site of a unit cell with the index $\mu=\mathrm{A}$ and the right site by $\mu=\mathrm{B}$, the only relevant correlations are thus
\begin{eqnarray*}
\langle \hat c_{m,\mu}^\dagger \hat c_{m,\mu}\rangle,\enspace \langle \hat c_{m,\mathrm{A}}^\dagger \hat c_{m,\mathrm{B}}\rangle,\enspace
\langle \hat c_{m,\mathrm{B}}^\dagger \hat c_{m+1,\mathrm{A}}\rangle,\enspace \langle \hat c_{m,\mathrm{A}}^\dagger \hat c_{m-1,\mathrm{B}}\rangle
\end{eqnarray*}
With this we find 
\begin{eqnarray}
H_\textrm{fict}&\approx& 2 \eta \sum_{l} \langle \hat c^\dagger_{l} \hat c_{l}\rangle \hat a_{l}^\dagger\hat a_{l} \label{eq:H-fict-approx}\\
&& + 2 \eta \sum_l \Bigl(\langle \hat c^\dagger_{l+1} \hat c_{l}\rangle \hat a_{l+1}^\dagger\hat a_{l} + h.a. \Bigr),\nonumber
\end{eqnarray}
where we have switched to a a simpler notation for the spatial indices $(m,\mathrm{A})\to l= 2m$ and $(m,\mathrm{B})\to l=2m+1$. Eq.\eqref{eq:H-fict-approx} is the
mean-field approximation  corresponding to a coupling Hamiltonian
\begin{eqnarray}
\tilde H_\eta &=&  2 \eta \sum_{l}  \hat c^\dagger_{l} \hat c_{l} \hat a_{l}^\dagger\hat a_{l} \label{eq:H-approx}\\
&&  +2 \eta \sum_l \Bigl( \hat c^\dagger_{l+1} \hat c_{l} \hat a_{l+1}^\dagger\hat a_{l} + h.a. \Bigr),\nonumber
\end{eqnarray}
which describes a \emph{local} density-density coupling and correlated \emph{nearest-neighbor} hopping.

\begin{figure}[htb]
	\begin{center}
	\includegraphics[width=0.95\columnwidth]{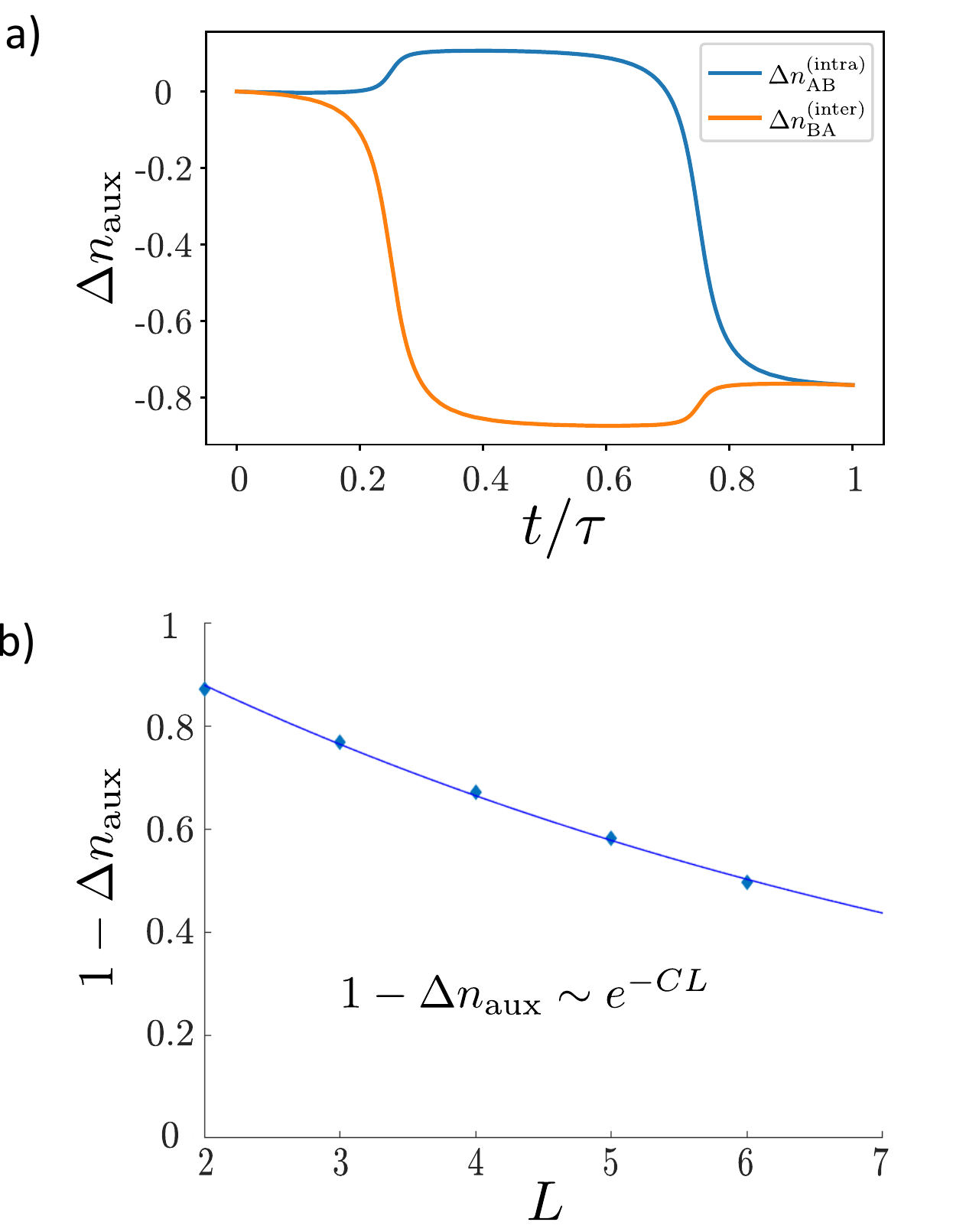}
	\end{center}
	\caption{(a) Particle transport in auxiliary system for system size of $L=2$ unit cells, and $\eta/\Delta_\mathrm{gap}^\mathrm{RM}= 0.25$ between sites within a unit cell (blue) and across unit cells (orange), using the approximate real-space coupling scheme
	\eqref{eq:H-approx}. All other parameter are as in Fig.\ref{fig:system}.
	(b) Deviation from quantized transport as function of 
	system size for strong coupling to auxiliary system $\eta/\Delta_\mathrm{gap}^\mathrm{RM}=1$. Even for this large coupling, clearly outside the perturbative regime,
	one recognizes  exponential decay with increasing system size $L$ and a quantized value of the transported charge is approached in the thermodynamic limit.}
	\label{fig:RM-finite}
\end{figure}

In Fig. \ref{fig:RM-finite}a we have shown numerical results for the adiabatic transport in the auxiliary system coupled to
a RMM chain by the approximate interaction \eqref{eq:H-approx} obtained from exact diagonalization for a system of $L=2$ unit cells (of 2$\times$2 sites). Shown is the time dependence  in one cycle of a Thouless
pump for repulsive interaction $(\eta>0)$ and $\eta/\Delta_\mathrm{gap}^\mathrm{RM}= 0.25$. One recognizes that the transport deviates from the expected value of unity. 
The deviation increases with increasing coupling strength $\eta$ but decreases exponentially with system size. This is shown 
to hold true even
for a very strong coupling $\eta/\Delta_\mathrm{gap}^\mathrm{RM}= 1$, i.e. outside of the perturbative regime, 
in Fig. \ref{fig:RM-finite}b.

\section{Topology transfer from interacting systems to auxiliary fermions}

We have seen that a chain of auxiliary fermions coupled to a one-dimensional lattice according to 
eq.\eqref{eq:H_eta} inherits the topological properties encoded in the ficticious Hamiltonian.
In particular  the transport in the auxiliary chain upon cyclic adiabatic variations of parameters directly detects 
the topological invariant of the ficticious Hamiltonian.

If the original system is in a gapped ground state of \emph{non-interacting} fermions, $\nu_\textrm{fict}$ is identical to
the  TKNN invariant.
It is near at hand to ask what happens in the presence of interactions? We will show in the following
that the  \emph{transfer of topology} also applies if the original system is an \emph{interacting} system
with an insulating man-body ground state. 

In the absence of interactions, gapped ground states
occur only at integer fillings of fermions per unit cell, which implies integer-valued topological charges. This changes with
interactions. Here gapped ground states can exist which have fractional fillings and the Lieb-Schulz-Mattis theorem
 \cite{Lieb-AnnPhys-1961} tells us that they are  degenerate \cite{Tao-PRB-1984}. 
We will thus have to discuss the topology transfer in the case of interactions with and without degeneracies separately.

\subsection{Non-degenerate ground state}


In Sec. \ref{free} we have shown that if the total system is prepared in a non-degenerate and gapped ground state $\vert \Phi_0\rangle=\vert\Phi_0^s\rangle\vert \Phi_0^a\rangle +{\cal O}(\eta)$, the charge transported in the
auxiliary chain is integer quantized and the integer is given by the winding number $\nu_\textrm{fict}$ of the Zak phase of the ficticious Hamiltonian.
%
%
Since  the discussion in Sec. \ref{sect-B} made no reference to the system being a non-interacting one,
all results apply equally to interacting particles. Thus 
%
\begin{equation}
Q_a = \nu_\textrm{fict}.\label{eq:Qa-int}
\end{equation}
%
%
Eq.\eqref{eq:Qa-int} gives us the license to interpret the winding number $\nu_\textrm{fict}$  as topological invariant also for
interacting one-dimensional systems with a non-degenerate, gapped ground state \cite{Rossini-PRB-2013}.

It is  near at hand to ask, if and what relation exists between $\nu_\textrm{fict}$ and e.g. the winding number $\nu_s$ of the many-body Zak phase $\phi_\textrm{Zak}^\textrm{MB}$ introduced by
Niu, Thouless and Wu (NTW) \cite{Niu-PRB-1985}
\begin{equation}
\phi_\textrm{Zak}^\textrm{MB}=  i \int_{-\pi/L}^{\pi/L}\!\!\! d\theta\,\, \bigl\langle  \Phi_0^s(\theta) \bigl\vert \partial_\theta\Phi_0^s(\theta)\bigr\rangle.\label{eq:MB-Zak}
\end{equation}
Here $\vert \Phi_0^s(\theta)\rangle$ is the non-degenerate many-body ground state of the twisted system Hamiltonian $H_s(\theta) = e^{-i\theta \hat X} H_s e^{i\theta \hat X}$, with $\hat X$ being the total position operator of all particles, and periodic boundary conditions are assumed. While for the examples we studied,
such as the superlattice Bose Hubbard model in a Mott insulating state with one boson per unit cell, it appeared that they are the same, 
i.e. $\nu_\textrm{fict}=\nu_s$,
a strict proof is not obvious. Furthermore a word of caution should be made: It is not guaranteed that the ficticious Hamiltonian remains gapped whenever the many-body ground state of the original system does. The latter holds true for non-interacting systems but may fail in the presence of interactions.


\subsection{Ground-state degeneracies and fractional transport}

Let us now discuss the case of a $d$-fold degenerate ground state $\vert \Phi_{0,\lambda}^s\rangle$ of the interacting system, with $\lambda=(1,2,\dots,d)$.
Furthermore let us restrict ourselves to cases where the degeneracy is accompanied by a spontaneous breaking of lattice-translational invariance. 
Then it is possible to find a basis $\{ \vert \Phi_{0,\lambda}^s\rangle\}$ such that the application of the lattice shift operator $\hat T$ by one unit cell
transforms between the basis states:
\begin{equation}
\vert \Phi_{0,\lambda+1}\rangle = \hat T\, \vert \Phi_{0,\lambda}\rangle.
\end{equation}
$d$-fold application of $\hat T$ returns any eigenstate back to itself (up to a phase). Real-space correlations $\langle \hat c_j^\dagger \hat c_{l}\rangle$  in any of the degenerate ground
states are then only invariant under $d$ successive lattice translations. It is therefore useful to introduce a new, enlarged unit cell. If the single-particle
Hamiltonian of the system has a unit cell of $p$ lattice sites, the enlarged unit cell consists of $p\cdot d$ sites, and the new Brillouin zone is correspondingly reduced.
The ficticious Hamiltonian expressed in this reduced Brillouin zone  has thus in general
$p\cdot d$ bands
\begin{eqnarray}
H_\textrm{fict}(t) = \eta \sum_k \sum_{\mu,\nu=1}^{p\cdot d}  \hat a_\mu^\dagger(k) \, {\sf m}_{\mu\nu}^\lambda(k,t) \,  \hat a_\nu(k)\label{eq:H-fict-deg}
\end{eqnarray}
where ${\sf m}_{\mu\nu}^\lambda(k,t)= \langle \Phi_{0,\lambda}(t)\vert \hat c_\mu^\dagger(k) \hat c_\nu(k)\vert \Phi_{0,\lambda}(t)\rangle$ is the covariance matrix
of single-particle correlations in the $\lambda$th ground state. In general the bands of $H_\textrm{fict}$ will be separated by gaps and the ficticious fermion system
will have a \emph{non-degenerate} gapped ground state when the number of auxiliary fermions is chosen appropriately.

The coupling Hamiltonian \eqref{eq:H_eta} that realizes the ficticious Hamiltonian for the auxiliary chain can also be rewritten in the reduzed Brillouin zone
\begin{eqnarray}
H_\eta = \eta \sum_k \sum_{\mu,\nu=1}^{p\cdot d}   \hat  c_\mu^\dagger(k) \hat c_\nu(k) \hat a_\mu^\dagger(k) \hat a_\nu(k).\label{eq:H_eta-deg}
\end{eqnarray}
Due to the degeneracy only a $d$-fold cycle of the 
parameters of the original system returns the ficticious Hamiltonian back to itself. Thus we expect that there is a quantized transport in the auxiliary system only
after $d$ pump cycles. This property is in full agreement with the corresponding property of the many-body Zak phase, eq.\eqref{eq:MB-Zak}. As shown by 
Niu, Thouless and Wu \cite{Niu-PRB-1985}, the topological invariant of an interacting system with a degenerate many-body ground state
is an integral of the Berry curvature  over an enlarged torus, extending  the
time integration to ${ \tau} \, d$
\begin{equation}
\nu_s = \frac{1}{2\pi} \int_0^{{\tau d}} \!\!\! dt   \int^{\pi/L}_{-\pi/L}\!\!\! d\theta \, \, \textrm{Im} \, 
\langle \partial_t \Phi_{0,\lambda}^s \vert \partial_\theta \Phi_{0,\lambda}^s\rangle.
\label{eq:nu-tot}
\end{equation}

\begin{figure}[htb]
	\begin{center}
	\includegraphics[width=0.98\columnwidth]{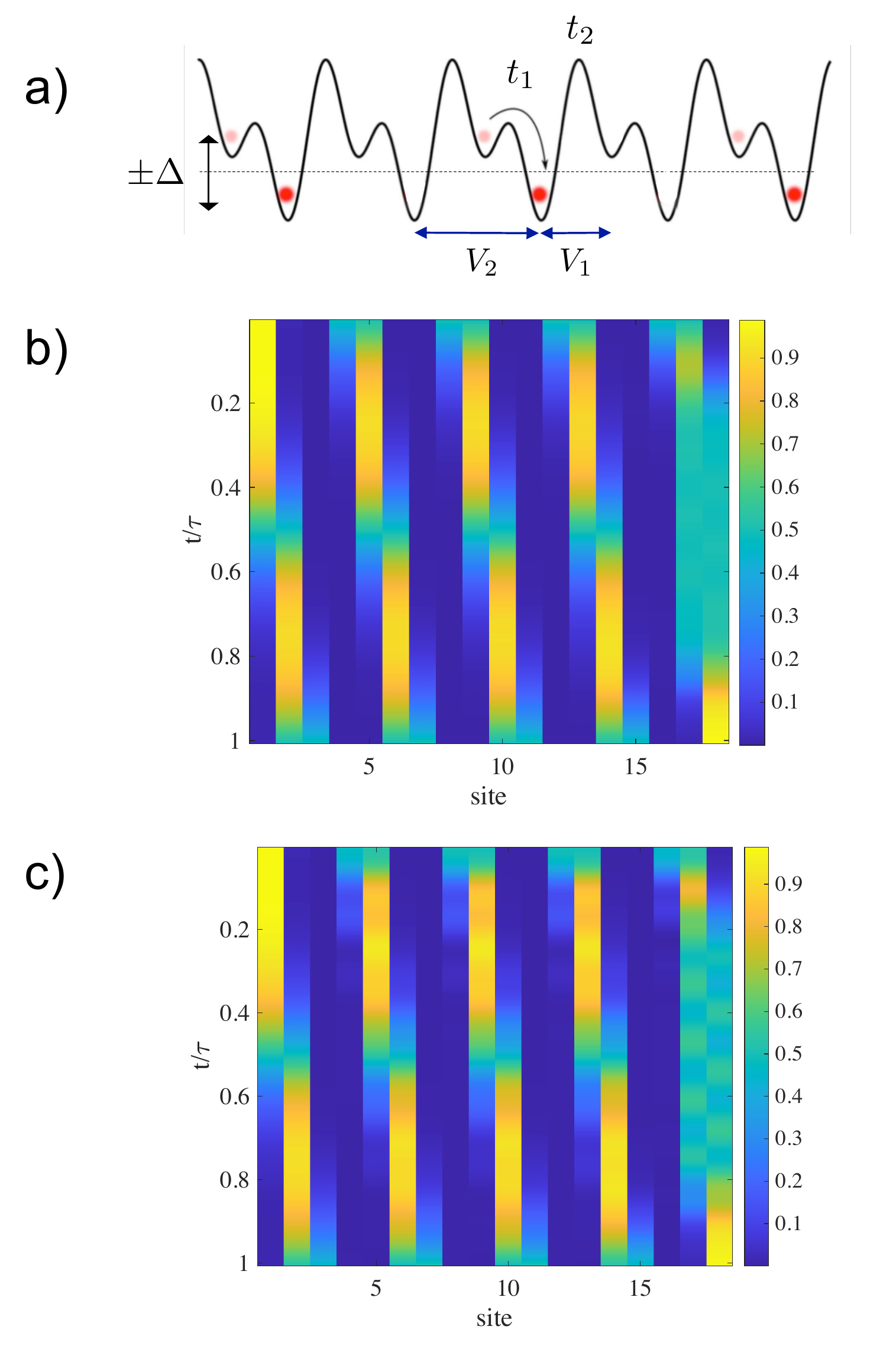}
	\end{center}
	\caption{(a) Extended super-lattice Bose Hubbard model with NN ($V_1$) and NNN ($V_2$) interactions. Particle transport at quarter filling in the bsoson system (b) and the auxiliary fermion system (b) for $\eta = 0.1$, $U=4000$, $V_1=2 V_2= 200$, and $t_{1,2}(t) = 0.5(1\pm \cos(2\pi t/\tau))$ and $\Delta(t)=\sin(2\pi t/\tau)$, $\tau=200$. }
	\label{fig:ExtSLBHM}
\end{figure}

Let us now consider as specific example with degenerate ground states, the one-dimensional extended superlattice Bose Hubbard model (ExtSLBHM)
\cite{ExtSLBHM,Zheng-PRB-2016}. The single-particle part of the Hamiltonian is identical to the RMM, eq.\eqref{eq:RMM}, only for bosons. In addition there are interactions between particles at the
same lattice site with strength $U$, nearest-neighbor (NN), and next-nearest neighbor couplings (NNN), $V_1$ and $V_2$, respectively,
see Fig. \ref{fig:ExtSLBHM}a. 
\begin{eqnarray}
H &=& - t_1\sum_{j,\textrm{even}} \hat a_j^\dagger \hat a_{j+1} - t_2\sum_{j,\textrm{odd}} \hat a_j^\dagger \hat a_{j+1} + c.c.\nonumber\\
&& + \Delta \sum_j (-1)^j \hat a_j^\dagger \hat a_j
+ \frac{U}{2} \sum_j \hat n_j(\hat n_j-1) \\
&& + V_1 \sum_j \hat n_j \hat n_{j+1}+ V_2 \sum_j \hat n_j \hat n_{j+2} \nonumber
\end{eqnarray}
where $\hat n_j=\hat a_j^\dagger \hat a_j$ denotes the particle number at lattice site $j$. 
For strong interactions this model has Mott-insulating phases at fractional fillings, e.g. at average filling $\rho_s=1/4$ per site. The corresponding
ground state is doubly degenerate and topologically non-trivial. Adiabatically varying the staggered potential $\Delta(t)$ and
the hoppings $t_{1}(t)-t_2(t)$ in a loop enclosing the origin realizes a topological Thouless pump. Since the ground state is
doubly degenerate, a single cycle transfers one ground state into the other one. Here two cycles are needed for an integer
quantized particle transport, which reflects the fractional topological charge of this model \cite{Zheng-PRB-2016,Li-PRB-2017}. If we prepare the system in one of the
two ground states $\vert \Phi_{0,\pm}\rangle$, with spontaneously broken translational invariance, the corresponding ficticious hamiltonian
${\sf m}_{\mu,\nu}^\pm$ has a single-particle gap, and an insulating many-body ground state exists at average
 filling of auxiliary fermions of $\rho_\textrm{aux}=1/4$ per lattice site. 

In order to calculate the particle transport in such a state from the full Hamiltonian, we use time-evolving block decimation (TEBD) simulations
\cite{TEBD,Daley-2004}, which are based on 
a representation of the many-body wavefunction in terms of matrix-product states (MPS) \cite{DMRG}. 
Since MPS simulations are much more difficult for periodic boundary conditions, we here choose a finite system of $2\times 18$ sites with open boundary conditions.
Furthermore we used the approximate real-space coupling Hamiltonian \eqref{eq:H-approx}.
The results of our simulations for strong interactions are shown in Fig. \ref{fig:ExtSLBHM}b and c, where the particle densities in the ExtSLBHM (b) and the auxiliary fermion chain (c) are shown as a function
of time during a single cycle of the Thouless pump. One recognizes that exactly $1/2$ particle was transported in both chains, reflecting the fractional topological charge of the ExtSLBHM.  We here have chosen $\eta < 0$ corresponding to an attractive interaction between the particles in the two chains. Note that due to the use of open boundary conditions we cannot simulate the full period of two pump cycles since then
a particle would be driven into excited states. Due to the open boundary conditions the ExSLBHM has an occupied edge state at the left end at $t=0$, where $\Delta =0$, and an occupied edge state at the right 
after a single cycle, i.e. at $t=\tau$. For the chosen parameter, which correspond to the atomic limit, the parallel transport in both chains becomes clearly visible.

%

\begin{figure}[htb]
	\begin{center}
	\includegraphics[width=0.8\columnwidth]{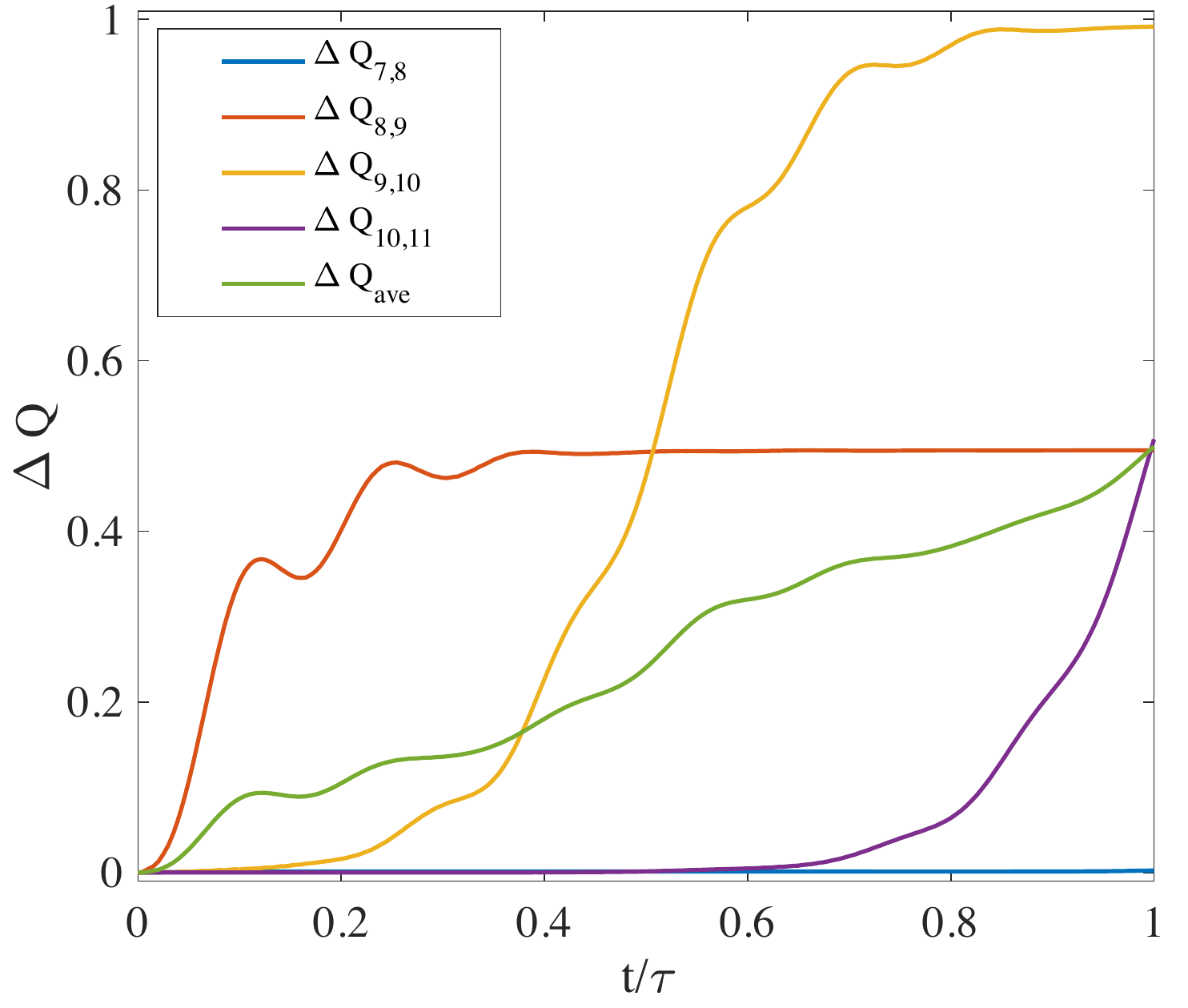}
	\end{center}
	\caption{Particle transport in the auxiliary system across different bonds between adjacent lattice sites as function of time during a single cycle of the Thouless pump
	for parameter of Fig.\ref{fig:ExtSLBHM}. While the net transport through different bonds varies due to the density-wave structure of the ground state, the average net transport
	$\Delta Q_\textrm{ave}$ is exactly $0.5$ reflecting the fractional topological charge of the ExSLBHM.}
	\label{fig:ExtSLBHM2}
\end{figure}

The time-evolution of the pump can be seen in more detail in Fig. \ref{fig:ExtSLBHM2}, where we have shown the particle transport in the auxiliary chain across different
bonds as function of time. One recognizes that the net transport averaged over one enlarged unit cell is exactly $0.5$ as expected.

%

\section{Outlook to finite-temperature systems}

\subsection{Ficticious Hamiltonian of a free-fermion system at $T>0$}

It is interesting to note that the discussion in Sec. \ref{sec:fict-H} for free fermions carries over to a finite-temperature state of the original system. 
As can be seen from eqs.\eqref{eq:m} and \eqref{eq:g} the
ficticious Hamiltonian remains gapped in thermal states as long as $T<\infty,$ i.e. $\beta >0$. 
For the eigenvalues $\tilde\epsilon_n^\textrm{fict}(k)$ in a thermal equilibrium state of (\ref{eq:H}) holds
\begin{equation}
\tilde\epsilon_n^\textrm{fict}(k) = \frac{\eta}{2}\left[ 1- \tanh\left(\frac{\beta(\epsilon_n(k)-\mu)}{2}\right)\right],
\end{equation}
i.e the spectrum is no longer flat for $T\ne 0$.
If the chemical potential $\mu$ is in the middle of a band gap of Hamiltonian (\ref{eq:H}) of size $\Delta$,  the 
ficticious Hamiltonian is also gapped:
\begin{equation}
\Delta_\textrm{gap}^\textrm{fict} = \vert \eta\vert \tanh\left(\frac{\beta \Delta}{4}\right).
\end{equation}
Increasing the temperature  leads to a reduction of the gap size, which 
approaches zero only at infinite temperature ($\beta =0$). 
Most importantly the  
Zak phase of the ficticious Hamiltonian has the same topological winding as that of the \emph{ground state} of the original Hamiltonian. 
From this it was concluded in Ref.\cite{Bardyn-PRX-2018} that non-interacting fermions at any finite temperature are topologically
equivalent to the ground state.

\begin{figure}[htb]
	\begin{center}
	\includegraphics[width=0.95\columnwidth]{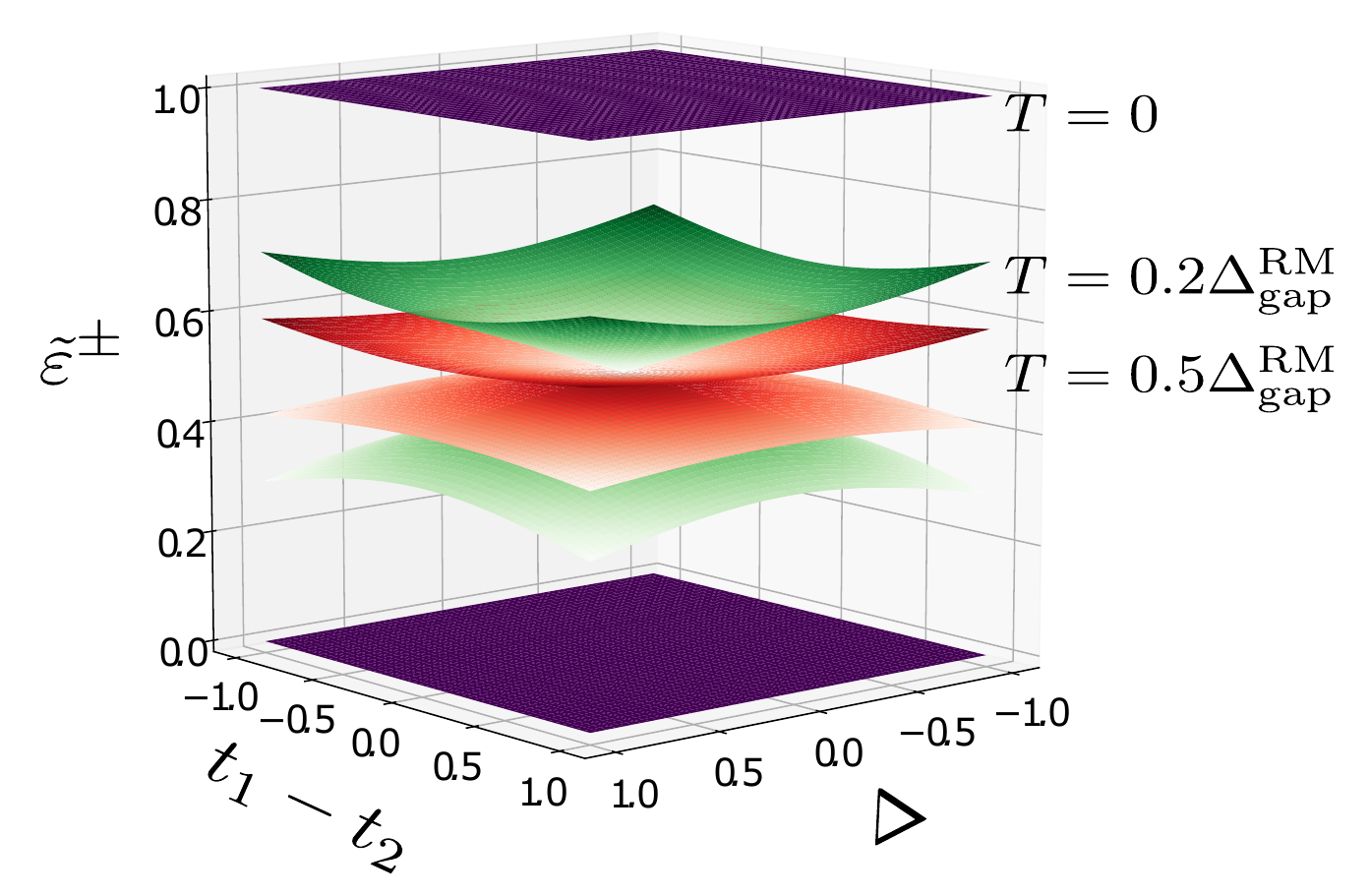}
	\end{center}
	\caption{Minimum energy of upper band $\tilde\varepsilon^+
	= \textrm{min}\bigl(\tilde\epsilon_0^+(k)\bigr)$ and maximum energy of lower band $\tilde\varepsilon^-=\textrm{max}\bigl(\tilde\epsilon_0^-(k)\bigr)$ of ficticious Hamiltonian of a Rice-Mele model
	at different temperatures $T$. Apart from the singular point of the original RMM ($\Delta=t_1-t_2=0$), the ficticious Hamiltonian remains
	gapped for all temperatures. (Note that for $T=0$ the bands are flat but there is a point singularity at the origin, which cannot be resolved in the graph.)}
	\label{fig:RM}
\end{figure}

Let us consider again the Rice-Mele model, eq.\eqref{eq:RMM}.
One easily calculates the finite-temperature covariance matrix, which has the following non-vanishing terms
\begin{eqnarray*}
		\langle \hat{c}_\mathrm{A}^\dagger(k) \hat{c}_\mathrm{A}(k)\rangle &=& \frac{1}{2} - \frac{\Delta}{2 \varepsilon_k} \tanh\left(\frac{\beta \varepsilon_k}{2}\right), \label{nA} \\
		\langle \hat{c}_\mathrm{B}^\dagger(k) \hat{c}_\mathrm{B}(k)\rangle &=& \frac{1}{2} + \frac{\Delta}{2 \varepsilon_k} \tanh\left(\frac{\beta \varepsilon_k}{2}\right), \label{nB} \\
		\langle \hat{c}^\dagger_\mathrm{A}(k) \hat{c}_\mathrm{B}(k) \rangle &=&  \frac{\left( t_1+t_2e^{ik}\right)}{2 \varepsilon_k}  
		\tanh\left(\frac{\beta \varepsilon_k}{2}\right), \\ 
		\langle \hat{c}^\dagger_\mathrm{B} (k) \hat{c}_\mathrm{A}(k)  \rangle &=&  \frac{\left( t_1+t_2e^{-ik}\right)}{2 \varepsilon_k}  
		\tanh\left(\frac{\beta \varepsilon_k}{2}\right). 
\end{eqnarray*}
This then leads to a ficticious Hamiltonian which, apart from an uninteresting overall energy shift, is again of the structure of the RMM with parameters directly related to those of the original model
\begin{eqnarray*}
		t_{1,2}\to \tilde{t}_{1,2}(k) &=& \frac{\eta}{2\varepsilon_k}
		\tanh\left(\frac{\beta \varepsilon_k}{2}\right) \,  t_{1,2}\\
		\Delta \to \tilde{\Delta}(k)  &=& \frac{\eta}{2\varepsilon_k}
		\tanh\left(\frac{\beta \varepsilon_k}{2}\right)
		\, \Delta 
\end{eqnarray*}
Its spectral gap is shown for different temperatures in Fig. \ref{fig:RM}. 
One recognizes that except for the singular point of the original RMM ($\Delta=t_1-t_2=0$) the ficticious Hamiltonian remains
gapped for all values of $T$. 

\subsection{Topological invariant of ficticious Hamiltonian: Many-body polarization}

We have argued in the previous sections that the topological properties of the \emph{ground state} of non-interacting lattice fermions are mapped to those
of the covariance matrix, respectively the ficticious Hamiltonian, and both are characterized by the same topological invariant. 
This invariant can be detected via a quantized transport in an auxiliary system using the topology-transfer scheme
discussed in this paper. 
In the case of an equilibrium state of the fermion system with \emph{non-zero temperature}, or even in a \emph{non-equilibrium steady state}, the ficticious Hamiltonian is still well defined
and one can ask for its topological properties encoded in the Zak phase (Wilson loop) of its lowest band. As was shown in \cite{Bardyn-PRX-2018} the latter can in fact be used
to generalize the concept of topology to finite-temperature and non-equilibrium steady states of non-interacting or Gaussian fermionic systems. It is interesting to note
that the topological invariant of the ficticious Hamiltonian for any pure or mixed Gaussian state is identical to the many-body polarization of the system, introduced by Resta \cite{Resta-PRL-1998,King-Smith-PRB-1993}
\begin{equation}
\phi_\textrm{Zak}^\textrm{fict}= 2\pi P \equiv \textrm{Im} \ln \textrm{Tr}\Bigl\{\rho \hat U\Bigr\}.
\end{equation}
Here $\hat U=e^{2\pi i \hat X}$ is the unitary momentum shift operator, which shifts the lattice momentum of every particle
by one unit. $\hat X = \frac{1}{L}\sum_{j=1}^L \sum_{s=1}^p (j + r_s) \hat n_{js}$ is the center of mass of all particles in the lattice consisting of $L$ unit cells, with the lattice constant set to one. $\hat n_{js}$ denotes the particle number in the $s$th site ($s\in \{1,2,\dots p\}$) of the $j$th unit cell and periodic boundary conditions are considered. $0\le r_s\le 1$ describes the position of the site within the unit cell. 

\section{summary and conclusion}

In the present paper we have shown that topological properties encoded in the covariance matrix of single-particle correlations of a one-dimensional
lattice system can be transferred to a second, auxiliary chain of non-interacting fermions, weakly coupled to the first. The coupling is constructed in such a way that
the auxiliary particles experience an effective mean-field Hamiltonian, called ficticious Hamiltonian, which is given by the covariance matrix of the first system. As a consequence 
an adiabatic cyclic variation of parameters of the original system induces a transport of auxiliary fermions in an insulating ground state. The charge pump is quantized and the transport is determined by the Zak-phase winding number of the ficticious Hamiltonian, which is an integer-valued topological invariant.
For non-interacting fermions this number is just the Thouless-Khomoto-Nightinghale-de Nijs (TKNN) invariant corresponding to either all occupied or all
unoccupied bands. We illustrated the topology transfer for a simple topologically nontrivial system of non-interacting fermions, the Rice-Mele model, coupled to an auxiliary
fermion chain, for which exact solutions for the state evolution can be derived. The coupling between the two chains, required for the topology transfer, is diagonal in momentum space and thus difficult to implement. We showed that it can however be well approximated by an interaction that contains only local density-density couplings and correlated nearest neighbor hoppings. 

In the presence of interactions in the original system, the transport induced in the auxiliary chain is still quantized and given by the winding number of
the ficticious Hamiltonian, which therefore defines a topological invariant for the interacting system. While without interactions, insulating states require integer fillings of fermions per unit cell, here gapped ground states can exists with fractional fillings and degeneracies. In such a case 
multiple loops in parameter space are needed for the eigenfunctions of the ficticious Hamiltonian to return to themselves, indicating fractional topological charges.
We illustrated this for the example of the extended superlattice Bose-Hubbard (ExtSLBH) model with nearest and next nearest neighbor interactions coupled to a chain of non-interacting fermions. The ExtSLBHM possesses a doubly degenerate Mott-insulating state at quarter filling of bosons. It has
a fractional topological charge of $1/2$ since only a two-fold  cyclic variation of the single-particle terms in the Hamiltonian leads to a winding of the many-body Zak phase 
by $2\pi$. Using numerical TEBD simulations we showed that a coupling of the ExtSLBHM to a chain of non-interacting fermions induces the same fractional transport in the auxiliary system. This suggests that the Zak-phase winding of the ficticious Hamiltonian is identical to the many-body topological invariant of Niu, Thouless and Wu (NTW).
While this is true in many cases, we could not derive a general relation to this invariant and such a relation may not exist in general.
Nevertheless if the winding number of the ficticious Hamiltonian  is non-trivial, it provides an observable invariant which allows to classify topological properties of an interacting system. 

The matrix describing the ficticious Hamiltonian of non-interacting fermions in thermal equilibrium is topologically equivalent to the corresponding matrix of the ground state. Thus, as sketched in the last section
of the paper,  the discussed transfer scheme may also provide a tool to directly observe topological invariants of finite-temperature states, such as the ensemble geometric phase in non-interacting \cite{Linzner-PRB-2016,Bardyn-PRX-2018} or interacting systems \cite{Unanyan-PRL-2020}.

\section*{Appendix}

Here we  give a proof that the mixed-term contribution in eq.(\ref{eq:Q1b}) vanishes. We consider
 \begin{eqnarray*}
    \varepsilon&=&\frac{i}{2\pi}\int_0^\tau \!\!\! dt\int_{-\pi/L}^{\pi/L} \!\!\!\! d\alpha\Big(\langle \partial_\alpha\Phi_0^{(0)}|\partial_t\tilde\Phi\rangle- \langle \partial_t\tilde\Phi|\partial_\alpha\Phi_0^{(0)}\rangle+\\
    &&\qquad\qquad\qquad+\langle \partial_\alpha\tilde\Phi|\partial_t\Phi_0^{(0)}\rangle-\langle\partial_t\Phi_0^{(0)}|\partial_\alpha\tilde\Phi\rangle\Big)
\end{eqnarray*}
Pulling out the derivatives in the bra vectors gives
 \begin{eqnarray*}
    \varepsilon&=&\frac{i}{2\pi}\int_0^\tau \!\!\! dt\int_{-\pi/L}^{\pi/L} \!\!\!\! d\alpha\Big(\partial_\alpha\langle \Phi_0^{(0)}|\partial_t\tilde\Phi\rangle- \partial_t\langle \tilde\Phi|\partial_\alpha\Phi_0^{(0)}\rangle+\\
    &&\qquad\qquad\qquad+ \partial_\alpha\langle\tilde\Phi|\partial_t\Phi_0^{(0)}\rangle-\partial_t\langle\Phi_0^{(0)}|\partial_\alpha\tilde\Phi\rangle\Big)\\
    &=& \frac{i}{2\pi} \int_0^\tau \!\!\! dt \Bigl(\langle \Phi_0^{(0)}|\partial_t\tilde\Phi\rangle + \langle\tilde\Phi|\partial_t\Phi_0^{(0)}\rangle\Bigr)^{\pi/L}_{\pi/L} +\\
    &&-\frac{i}{2\pi} \int_{-\pi/L}^{\pi/L} \!\!\!\! d\alpha \Bigl(\langle \tilde\Phi |\partial_\alpha\Phi_0^{(0)}\rangle+\langle\Phi_0^{(0)}|\partial_\alpha\tilde\Phi\rangle\Bigr)^{\tau}_{0}
\end{eqnarray*}
Making use of the orthogonality of the state vectors $\langle \Phi_0^{(0)}\vert \tilde \Phi\rangle =0$ one thus finds
 \begin{eqnarray*}
    \varepsilon &=& \frac{i}{2\pi} \int_0^\tau \!\!\! dt \Bigl(\langle \Phi_0^{(0)}|\partial_t\tilde\Phi\rangle -c.c.\Bigr)^{\pi/L}_{-\pi/L} +\\
    &&-\frac{i}{2\pi} \int_{-\pi/L}^{\pi/L} \!\!\!\! d\alpha \Bigl(\langle \tilde\Phi |\partial_\alpha\Phi_0^{(0)}\rangle-c.c.\Bigr)^{\tau}_{0}
\end{eqnarray*}
Since upon a full cycle in either $t$ or $\alpha$ the perturbed ground state $\vert \Phi_0^{(1)}\rangle$ has to return to itself apart from a phase factor, one has
\begin{eqnarray*}
\vert \Phi_0^{(0)}(\alpha, \tau)\rangle &=& e^{i\vartheta(\alpha)} \vert \Phi_0^{(0)}(\alpha, 0)\rangle,\\
\vert \tilde \Phi(\alpha, \tau)\rangle &=& e^{i\vartheta(\alpha)} \vert \tilde\Phi(\alpha, 0)\rangle,\\
\vert \Phi_0^{(0)}(\pi/L, t)\rangle &=& e^{i\lambda(t)} \vert \Phi_0^{(0)}(-\pi/L, t)\rangle,\\
\vert \tilde \Phi(\pi/L, t)\rangle &=& e^{i\lambda(t)} \vert \tilde\Phi(-\pi/L, t)\rangle.
\end{eqnarray*}
which yields 
\begin{eqnarray*}
\langle \Phi_0^{(0)}|\partial_t\tilde\Phi\rangle\Bigr\vert_{\frac{\pi}{L}} &=& i\frac{\partial \lambda(t)}{\partial t} \langle \Phi_0^{(0)}|\tilde\Phi\rangle\Bigr\vert_{\frac{\pi}{L}}
+\langle \Phi_0^{(0)}|\partial_t\tilde\Phi\rangle\Bigr\vert_{-\frac{\pi}{L}}\\
&=&\langle \Phi_0^{(0)}|\partial_t\tilde\Phi\rangle\Bigr\vert_{-\frac{\pi}{L}},\qquad\textrm{etc.}
\end{eqnarray*}
where in the second line we have used again the orthogonality $\langle \Phi_0^{(0)}\vert \tilde \Phi\rangle =0$.
This finally gives
\begin{equation}
\varepsilon=0
\end{equation}

\subsection*{acknowledgement}
The authors gratefully acknowledge financial support from the DFG through SFB TR 185, project number
277625399 as well as valuable input from D. Linzner in the initial stages of the project.




\end{document}